# Violation of Bohr's Complementarity: One Slit or Both?

Shahriar S. Afshar[1]

*Physics Department, Rowan University, Glassboro, NJ 08028*

**Abstract.** We have implemented a novel double-slit "which-way" experiment which raises interesting questions of interpretation. Coherent laser light is passed through a converging lens and then through a dual pinhole producing two beams crossing over at the focal point of the lens, and fully separating further downstream providing which-way information. A thin wire is then placed at a minimum of the interference pattern formed at the cross-over region. No significant reduction in the total flux or resolution of the separated beams is found, providing evidence for coexistence of perfect interference and which-way information in the same experiment, contrary to the common readings of Bohr's principle of complementarity. This result further supports the conclusions of the original experiment by the author in which an imaging lens was employed to obtain which-way information. Finally, a short discussion of the novel nonperturbative measurement technique for ensemble properties is offered.

## INTRODUCTION

Wave-particle duality as embodied in Bohr's Principle of Complementarity (PC) has been at the heart of quantum mechanics since its inception [1]. The celebrated Bohr-Einstein debate revolved around this issue and was the starting point for many of the illuminating experiments conducted during the past few decades [2]. In the context of double-slit "which-way" experiments, PC dictates that "The observation of an interference pattern and the acquisition of which-way information are mutually exclusive" [3]. The current literature and textbooks [3,4], state that in a particular experiment, we can either have (i) perfect fringe visibility and no which-way information, or (ii) full which-way information and no fringes. While the above two criteria successfully hold when we apply direct (perturbative) measurements of the photons, indirect measurements of the interference pattern or which-way detection have thus far been ignored by investigators. Here we report a simple modified double-slit experiment which shows evidence for presence of highly visible interference and very reliable which-way information in the same experimental arrangement for the same photons. This experiment is a follow-up to the original experiment by the author in which an imaging lens was used to obtain which-way information [5,6]. The improvement here is the transparency of reliability of which-way information by virtue of the law of conservation of linear momentum.

In 1979 Wheeler suggested an experiment [7] shown in Fig. 1, which can readily demonstrate the mutual exclusivity of criteria (i) and (ii) *when conventional measurement schemes* are applied thus confirming the validity of PC ins such experiments. Coherent light is incident on a converging lens immediately after which dual pinholes (double-slits in Wheeler's original schematics) are placed. The resulting diffracted beams are apodized by passing through an aperture stop (AS) which allows only the Airy disk of each single pinhole diffraction pattern and blocks the higher order diffraction rings. By placing a photosensitive device at $\sigma_1$, the focal plane of the lens where the two beams cross over, we can build-up an interference pattern with a very high fringe visibility. Clearly, this observation does not provide any which-way information, and thus confirms criterion (i).

---

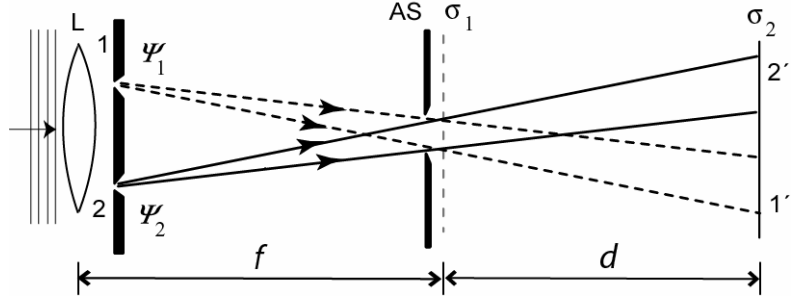

**FIGURE 1**. Wheeler's experiment to demonstrate complementary wave and particle aspects of light. By placing a photosensitive surface at plane $\sigma_1$, an interference pattern can be observed. Alternatively, removing the detector at $\sigma_1$ allows unitary evolution, and spatial separation of the two beams farther downstream at plane $\sigma_2$ thus providing which-way information by virtue of the law of conservation of linear momentum. Note however, that now no evidence exists throughout the experiment of interference, seemingly confirming the validity of Complementarity.

The spatial photon probability distribution at $\sigma_1$ can be represented by the following equation:

$$P_{12} = |\psi_1|^2 + |\psi_2|^2 + \Gamma , \qquad (1)$$

where $\psi_1$ and $\psi_2$ are the wavefunctions originating from pinholes 1 and 2 respectively, and $\Gamma = \psi_1^* \psi_2 + \psi_1 \psi_2^*$ is the usual interference cross terms.

Alternatively, we can remove the photosensitive surface at $\sigma_1$, and allow the beams to evolve unitarily and become well separated spatially in order to obtain highly reliable which-way information by observing the well-resolved beam profiles $1'$ and $2'$ of the corresponding pinholes 1 and 2 at the image plane $\sigma_2$. Although the images are not perfect representations of the pinholes due to the finite aperture size of the lens, the "bleeding" from one pinhole to the other beam can be made negligible by ensuring that the central peaks of beam profiles are angularly and spatially well-separated. *This can be achieved by ensuring that the diameter of the Airy disks at $\sigma_1$ is smaller than the center-to-center separation of the dual pinholes.* Notably, this requirement leads to a very small IP. The measurement difficulties associated with such a small pattern produce lower statistical accuracies discussed later.

Assuming $\psi'_1$ and $\psi'_2$ are the unitarily evolved $\psi_1$ and $\psi_2$ wavefunctions mapped onto plane $\sigma_2$, the probability distribution there can be written as:

$$P'_{12} = |\psi'_1|^2 + |\psi'_2|^2 + \Gamma' . \qquad (2)$$

In contrast to Eq. (1), since $\psi'_1$ and $\psi'_2$ have negligible overlap at $\sigma_2$, there is essentially no interference term on the RHS of Eq. (2), i.e. $\Gamma' = \psi'^*_1 \psi'_2 + \psi'_1 \psi'^*_2 \approx 0$. Therefore, at $\sigma_2$ we have reliable which-way information, but no interference. At first glance, it seems that this result is in agreement with criterion (ii). However, the question we attempt to answer in this paper is whether the interference pattern at $\sigma_1$ is destroyed by the *later* act of obtaining which-way information at $\sigma_2$, as Wheeler claims [7].

## IMPLEMENTATION OF CONVENTIONAL MEASUREMENTS

In the first experiment we duplicated Wheeler's original setup and obtained the complementary measurements at $\sigma_1$ and $\sigma_2$ *by conventional means*, i.e. directly observing the incident flux. We used a class II battery-powered Hotech model LMH-CP1-650A diode laser operating at $\lambda = 650$ nm, with an optical output power of roughly 1mW as the source. The collimated laser beam illuminated a converging lens with $f = 20$ cm immediately after which a dual-pinhole (obtained as a custom part from National Apertures Inc., Salem, NH) was placed. The dual pinhole was mounted such that the line connecting the centers of the pinholes was horizontal. Each pinhole had a diameter of 250 μm and they were separated by a distance of 2 mm ($> 3000\lambda$) between centers. The dual pinhole was mounted in a two-degree-of-freedom mount that allowed (1) horizontal translation (to ensure equal flux through each pinhole) and (2) rotation about the axis of beam propagation. The dual pinhole was constructed from 12.7 micron thick flat stainless steel, non-magnetic blackened material to minimize stray and scattered light. We will describe the rest of the optical system in a right handed coordinate system where $+x$ is horizontal, $+y$ is vertically up, and $+z$ is along the propagating beam. The origin of the coordinate system is taken to be the midpoint between the pinholes.

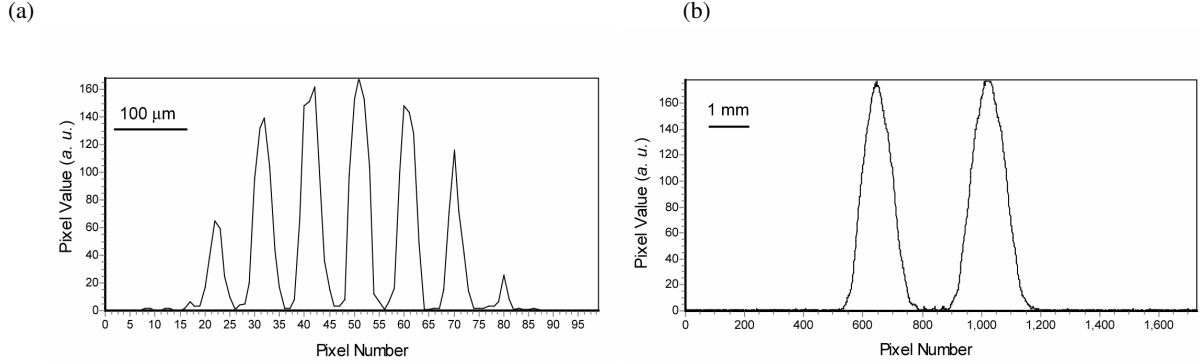

**FIGURE 2.** Direct measurement of the complementary wave and particle aspects of light. (a) The interference pattern observed at $\sigma_1$. (b) The intensity profile of the images at $\sigma_2$. The beam profiles $1'$ and $2'$ are clearly well-resolved. Light intensity I is measured in arbitrary units (*a.u.*) of graylevel output of the CCD. Note that these two observations are mutually-exclusive for the same photons.

A precision circular aperture of diameter 3 mm was placed at $z = 13$ mm, in a mount that allowed us to transmit light from either of the pinholes, or from both, depending upon its horizontal position. The diffracted beam from the pinholes was apodized by passing through a precision circular AS of diameter 500 μm, located at $z = 210$ mm. The center of this aperture was carefully aligned with the z axis of the system.

The plane $\sigma_1$ was located at $z = f = 20$ cm, where we chose to directly interrogate the interference pattern (IP) from the dual pinholes. Figure 2(a) depicts the intensity profile of the IP measured at $\sigma_1$. Using the equation for fringe visibility

$$V = (I_{max} - I_{min})/(I_{max} + I_{min}) , \qquad (3)$$

where $I_{max}$ is the maximum intensity of a bright fringe and $I_{min}$ is the minimum intensity of the adjacent dark fringe, we obtain $V= 0.99 \pm 0.01$ for the central fringes. The plane $\sigma_2$, was located at $z = 51.5$ cm ($d = 32.5$ cm) where the which-way information was obtained. Fig. 2(b) shows the intensity profile of the well-resolved beams. Less than $10^{-6}$ of the peak value irradiance from either image is found to enter the other channel, ensuring virtually 100% reliability for the which-way information at $\sigma_2$.

It is evident from the direct observational data in Fig. 2(a) that the photons are prepared in a perfectly coherent state and the initial conditions for the interference of the two wavefunctions originating at each pinhole are met. This cumulative data is the result of allowing the wavefucntion of each single photon (the superposition of $\psi_1$ and $\psi_2$) arriving at the detector placed in $\sigma_1$ to collapse into a singular spatial coordinate on the 2D detector, the cumulative result of which is the IP. However, the question remains as to the "reality" of interference at $\sigma_1$ when the collapse of the wavefunctions takes place at $\sigma_2$ where no interference can be observed.

The law of conservation of linear momentum compels us to accept that a photon in a particular bright spot on $\sigma_2$ must have originated from the corresponding pinhole. This line of reasoning (which must be upheld if we are not to forgo the conservation laws) leads one to conclude that in line with PC, we do not have interference at $\sigma_1$, as each photon can be assigned to only one pinhole. Paradoxically however, we shall see in the next section that this argument fails, due to the fact that contrary to common belief, *we can indeed gain information about the presence of interference at $\sigma_1$ without destroying (in principle) any photons, or disturbing the wavefunctions sufficiently enough to destroy the which-way information.*

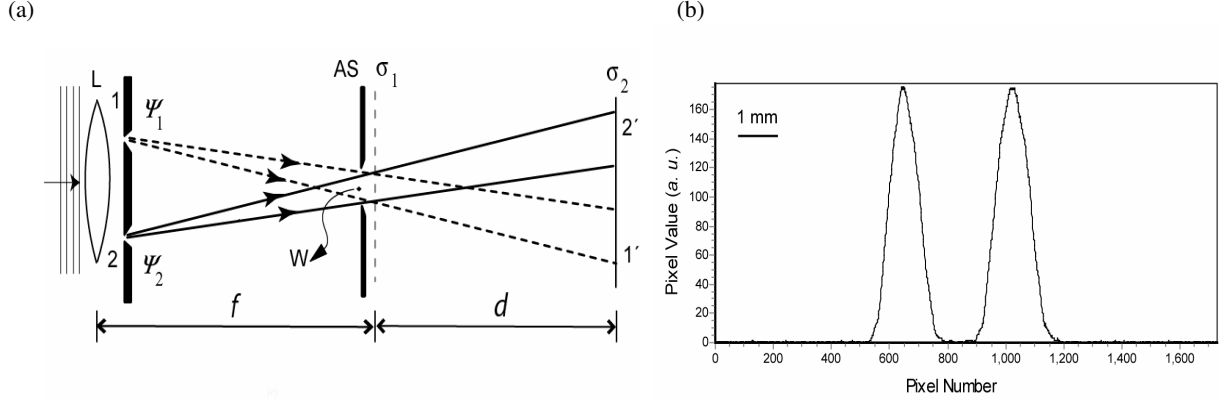

**FIGURE 3.** The modified crossed-beam "which-way" (a) experimental setup and (b) beam profiles at $\sigma_2$.

# IMPLEMENTATION OF NON-PURTURBATIVE MEASUREMENT

So much for the orthodox measurement techniques! To test Wheeler's (and by proxy Bohr's) claim regarding the destruction of IP at $\sigma_1$, we introduced a vertical wire mask W at $\sigma_1$, with the wire thickness of 10 μm placed at the central dark fringe of the IP as shown in Fig. 3(a). The resolution of the beam profiles at $\sigma_2$ was then measured and *was found to be negligibly affected by the presence of the wire* in front of the lens as shown in Fig. 3(b). This is not surprising, when we realize that the only means by which the resolution of the beam profiles could be reduced is the scattering of incident light by the wires. Since no light is incident on the wire, no scattering takes place, and the result is *essentially* similar to the case when there was no wire(s) present. A comparison between the image data in Fig. 2(b) and Fig. 3(b) clearly demonstrates the negligible effect of the wires on the beam profiles.

In addition to the analysis of the resolution of beams, another measure of the effect of the wire is the amount of light blocked by the wire(s). The result of the reduction in total radiant flux relative to the control (when no wire is present) is shown in Fig. 4(a-c). The CCD used for these measurements was a Canon 20D, with shutter speed (integration time) of 1/2000 s and ISO of 100.

Fig. 4(a-c) depicts the control runs (left), where no wire is present, the beam profiles (middle) and the runs in which the wire is present (right). $\Phi_C$, the total radiant flux of control runs in the Airy disks, is used to normalize the measurements in the experiments. The normalized reduction in the total radiant flux (in percent) due to the presence of the WG in the subsequent runs is defined as $R = 100\,(\Phi_C - \Phi_{Observed})/\Phi_C$, where $\Phi_{Observed}$ is the total flux received in the Airy disks of each run with wire present.

Fig. 4(a) shows the case in which pinhole 1 is closed. In this configuration, no IP can form at $\sigma_1$, therefore there would be incident photons on the wire, which in turn, attenuates and diffracts the transmitted light gathered by the detector. The loss of the radiant flux due to the wire in this case is theoretically calculated to be 2% of $\Phi_C$. The normalized amount of light absorbed by the wires is found to be $R = (2.1 \pm 0.5)\%$ by the analysis of the data, which matches the above theoretical value very well. Also, as expected, it is evident from the density plot of the $D_2$ output that the point-spread function of beam profile $2'$ has been significantly degraded in comparison to that of the control case. Fig. 4(b) shows the same results for the case when pinhole 2 is closed.

Fig. 4(c) depicts the result for the case when both pinholes are open and for the control and for the case when the wire is present. In this case, we find $R = (0.3 \pm 0.2)\%$. Clearly, no significant reduction in total radiant flux due to the wire is found. Also, the flux distribution is much less disturbed in comparison to Fig. 4(a and b).

According to PC the results of the last two experiments shown in Fig. 4(c) must be significantly different, if the IP at $\sigma_1$ is to be destroyed due to the which-way measurement at $\sigma_2$. But we find that there is no statistically significant difference between the two results, regardless of the lack or presence of the wire(s).

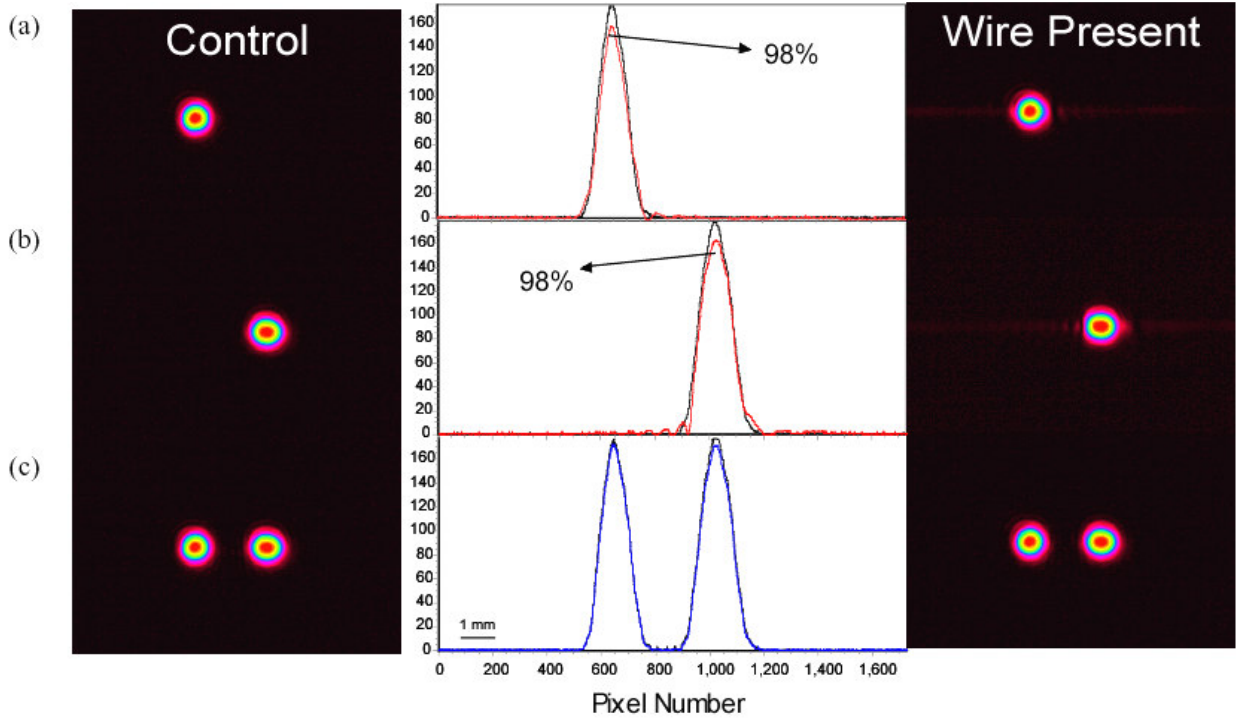

**FIGURE 4**. Results of six different experimental setups. In each of the three sets, the profiles of a pair of results are shown together for easier comparison of the flux distribution. (a) Beam profile obtained at $\sigma_2$ when pinhole 1 is closed for both cases in which wires are present and the one where there is no wire (control). (b) Same as (a) but now pinhole 2 is closed. (c) Results for the cases when both pinholes are open. Note that these two profiles overlap tightly.

## CONCLUSION

The results of this experiment confirm the earlier findings by the author [5, 6]. We have shown that we can establish the presence of perfect interference without appreciably disturbing or attenuating the interfering wavefunctions. The null measurement achieved by the passive presence of the wire(s) demonstrates for the first time that one *can* make meaningful measurements *without* an interaction or quantum entanglement with the measuring device i.e. the wire(s). This observation necessitates a revision of the current theory of measurement in which a measurement *always* leads to a change in the quantum state of the detector, which will be fully addressed elsewhere [8].

These results also highlight the inadequacy of classical language of waves and particles in describing seemingly simple experiments, for if we insist on using the wave *picture* to describe the lack of reduction of radiant flux and beam profile resolution by the wire(s), then we are forced to describe the pattern observed at plane $\sigma_2$ as an interference pattern *without* any fringes as evidence of the interference. **While it is true that PC still holds for perturbative methods of measurement, which involve which-way markers, entanglement, and direct measurements, indirect measurement of ensemble properties such as interference, as achieved in this experiment, provides evidence for the coexistence of complementary wave and particle behaviors in the same experimental setup.** If we (erroneously) insist on using the language used by Bohr and Einstein in their debates, then we would have to conclude that the photons in our last experiment, in fact went through both pinholes, and yet simultaneously, through one or the other: a logical impossibility! That said, it is hard to envision a common mode of language that best describes the results of this experiment, without an appeal to mathematical formalism.

The results of this experiment can be improved upon by the introduction of multiple wires mask. We also predict similar results for single-photons and other quanta in analogous experiments.

## ACKNOWLEDGEMENTS

The author thanks G. B. Davis, D. W. Glazer, J. Grantham and other financial supporters of this research. The author also thanks Rowan University Department of Physics for their hospitality and support.